\documentclass[fleqn,11pt]{wlscirep}
\usepackage{bm}
\usepackage{graphicx}
\usepackage{color}
\usepackage[normalem]{ulem}
\usepackage{epstopdf}
\usepackage{amsmath}
\usepackage{amssymb}
\usepackage{soul}
\usepackage{color}
\usepackage{enumitem}

\def\ep{\varepsilon}

\title{Driven nonlinear nanomechanical resonators as digital signal detectors}

\author[1,*]{Yukihiro Tadokoro}
\author[1]{Hiroya Tanaka}
\author[2]{M. I. Dykman}
\affil[1]{Toyota Central R\&D Labs., Inc., Nagakute, Aichi 480-1192, JAPAN}
\affil[2]{Michigan State University, East Lansing, MI 48824, USA}
\affil[*]{tadokoro@mosk.tytlabs.co.jp}

\begin{abstract}
Because of their nonlinearity, vibrational modes of resonantly driven nanomechanical systems have coexisting stable states of forced vibrations in a certain range of the amplitude of the driving force. Depending on its phase, which encodes binary information, a signal at the same frequency increases or decreases the force amplitude. The resulting force amplitude can be outside the range of bistability. The values of the mode amplitude differ significantly on the opposite sides of the bistability region. Therefore the mode amplitude is very sensitive to the  signal phase. This suggests using a driven mode as a bi-directional bifurcation amplifier, which switches in the opposite directions depending on the signal phase and provides an essentially  digital output. We study the operation of the amplifier near the critical point where the width of the bistability region goes to zero and thus the threshold of the signal amplitude is low. We also develop an analytical technique and study the error rate near the threshold. The results apply to a broad range of currently studied systems and extend to micromechanical systems and nonlinear electromagnetic cavities.
\end{abstract}
\begin{document}

\flushbottom
\maketitle
%
%
\thispagestyle{empty}

\section*{Introduction}

Nano- and micro-electro-mechanical systems (NEMS and MEMS) provide a promising platform for mass, charge, and force detection with extremely high sensitivity \cite{Sazonova2004,Rugar2004,Aldridge2005,Buks2006,Bunch2007,Jensen2008,DeMartini2008,Lee2010,Eom2011,Chaste2012,Hong2012,Puller2013,Moser2013,Venstra2013,Singh2014,Hanay2015,Fong2015,Tao2016}.  An important and well-established feature immediately related to the smallness of NEMS and MEMS is the nonlinearity of their vibrational modes \cite{Schmid2016}. It leads to the broadening of the spectra of the response to an external field \cite{DK_review84}, which is not related to energy dissipation, and can limit the performance of NEMS- and MEMS-based devices \cite{Yang2016}.

\begin{figure}[h]
\begin{center}
\includegraphics[height=3.2cm]{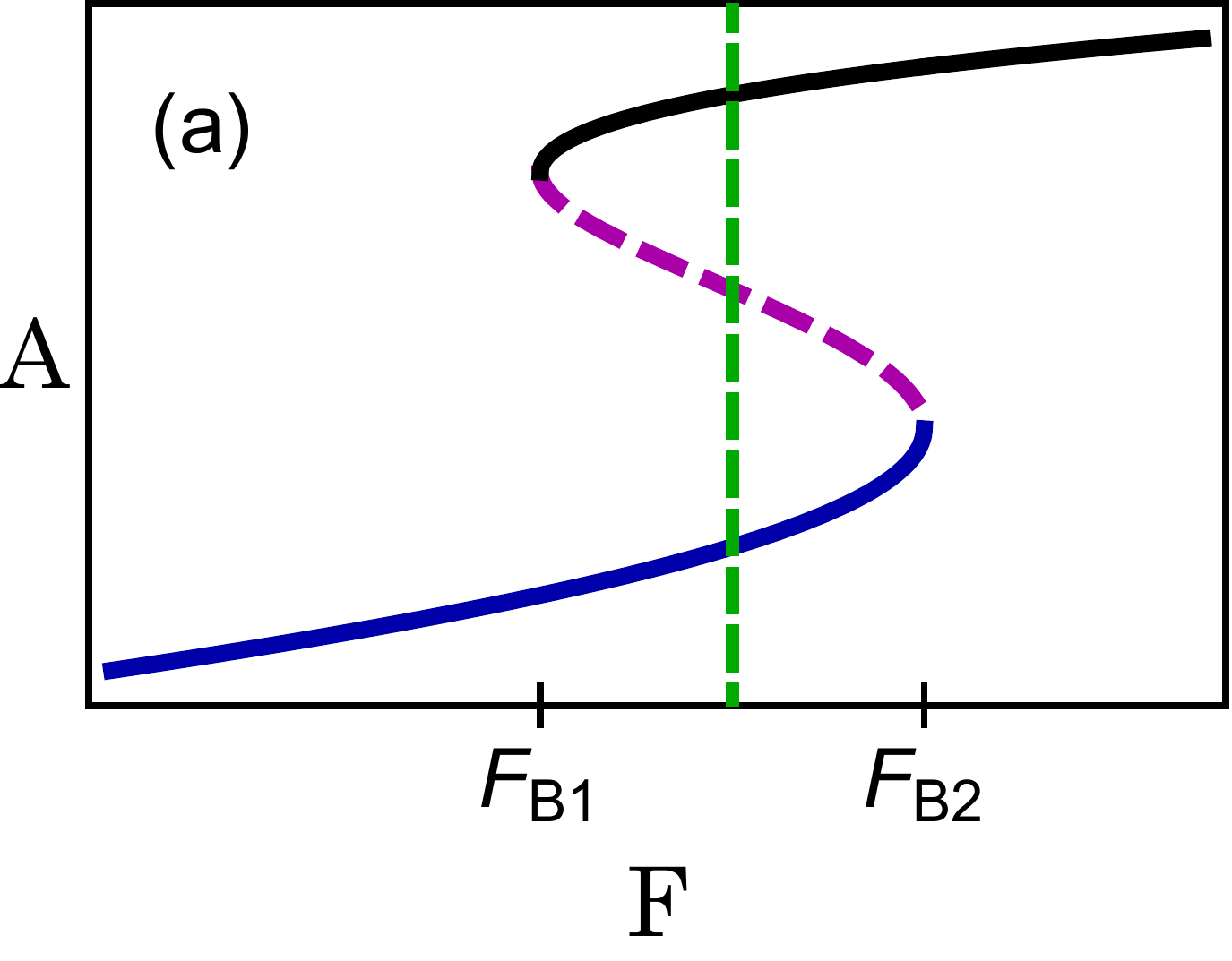} \hspace{0.15in}
\includegraphics[height=3.2cm]{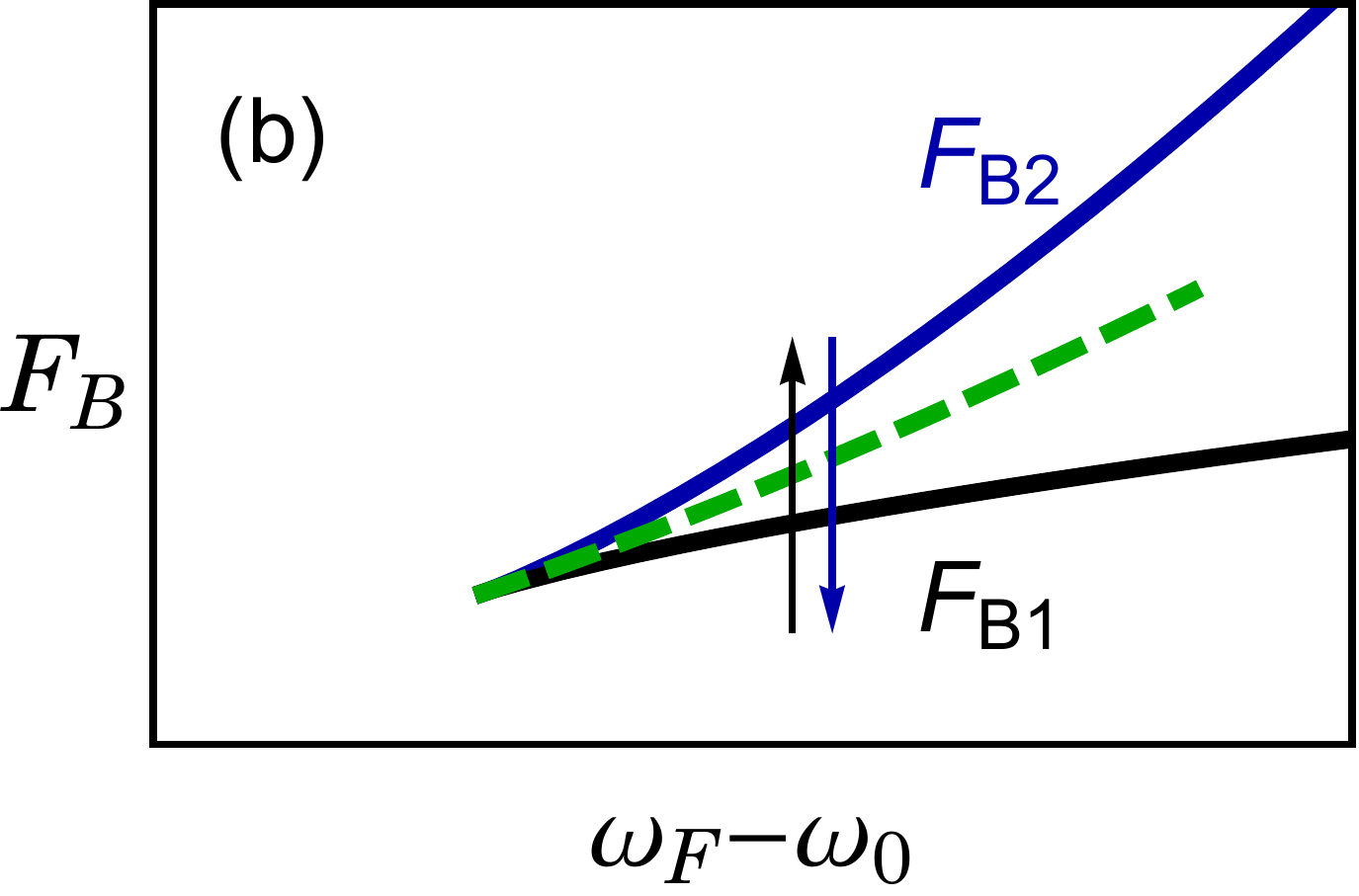}\hspace*{0.15in}
\includegraphics[height=3.1cm]{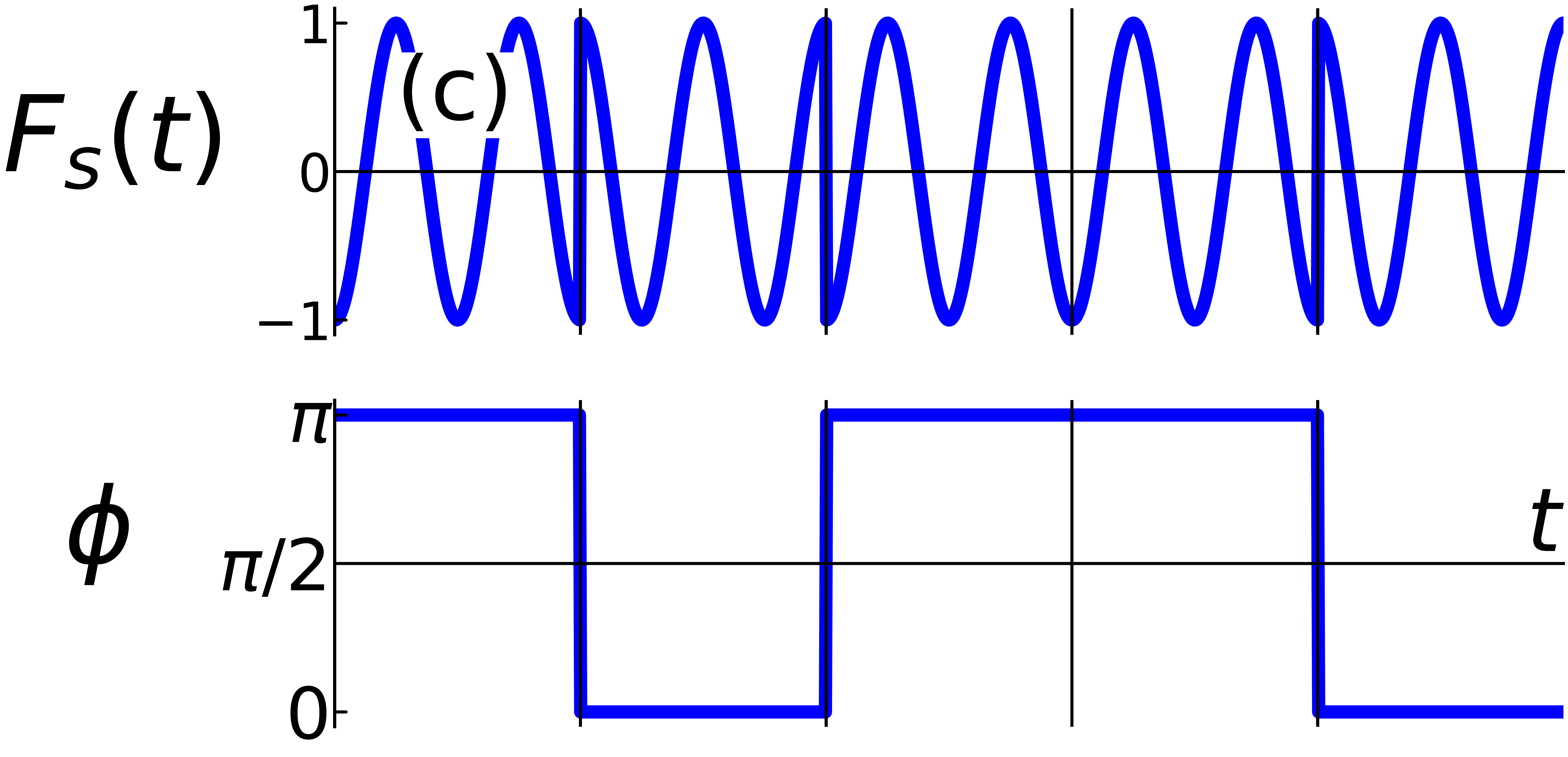}\hspace*{0.1in}
\end{center}
\caption{{\bf Driven nonlinear mode as a bi-directional bifurcation amplifier} (a) The amplitude of forced vibrations $A$ of a nonlinear mode as a function of the amplitude $F$ of the driving resonant force. Solid lines show stable values of $A$, dashed magenta line shows the unstable stationary value of $A$; $F_{B1}$ and $F_{B2}$ are the bifurcational values of $F$ where one of the stable branches ends. The vertical line indicates the optimal value $(F_{B1} + F_{B2})/2$ of $F$ in the absence of a signal. (b) The dependence of the bifurcation amplitudes $F_{B1,2}$ on the drive frequency near the critical point where the $F_{B1}$ and $F_{B2}$ merge. The dashed line shows $(F_{B1}+F_{B2})/2$. In the presence of the signal, the total drive amplitude $F\pm F_S$ is either smaller than $F_{B1}$ or larger than $F_{B2}$ and switches between its values depending on the phase of the signal. This switching is indicated by the arrows. It leads to a large change of the vibration amplitude, as seen from panel (a). (c) The signal is encoded in the phase $\phi$ of the periodic force $F_S(t)$ (the lower panel). The upper panel is a sketch of the change of $F_S(t)$ as $\phi$ changes from $0$ to $\pi$ and back. We consider the case where the number of the oscillation periods between the phase switching is large. }
\label{fig:hysteresis}
\end{figure}

A consequence of the mode nonlinearity is that, when a mode is driven close to resonance, the vibrations display bistability: for the same value of the driving force amplitude $F$, the vibration amplitude $A$ can take on one of the two values. For weakly damped modes, the bistability occurs already for a comparatively weak driving. The dependence of the vibration amplitude on the drive amplitude is illustrated in Fig.~\ref{fig:hysteresis}(a) \cite{LL_Mechanics2004}; the weak-driving bistability is manifested for the drive amplitude limited by the two values, the bifurcation points $F_{B1}$ and $F_{B2}$.  In recent years, this  bistability has found an important application in bifurcation amplifiers used in quantum measurements \cite{Vijay2009,Murch2012,Ithier2014}. The underlying idea of such an amplifier is that the input signal modifies the parameters of the system in such a way that the bifurcation point is crossed; for example, one can think of changing the drive amplitude $F$ so that it crosses point $F_{B1}$ or $F_{B2} $ in Fig.~\ref{fig:hysteresis}(a). As a result, the amplitude of the vibrations changes very strongly, i.e., the system strongly and nonlinearly amplifies the input signal. The bistability of nonlinear vibrations was also suggested as a means of measuring, with high sensitivity, the resonant frequency and nonlinearity parameters of the vibrations \cite{Aldridge2005}.

In the present paper we propose and study a qualitatively new application of nonlinear NEMS and MEMS: to use the bistability of their forced vibrations for signal processing. The signal frequencies in communication systems lie in the broad range from 1 MHz  to 10 GHz. However, a significant portion of the currently used communication devices operate at the lower end of this range, with frequencies, $\lesssim 100$~MHz. The characteristic frequencies of NEMS and MEMS are usually also $\lesssim 100$~MHz, although recently high-quality mechanical resonators operating at a few-GHz frequencies have been also developed \cite{O'Connell2010,Chu2017}. This shows that there is a frequency match between NEMS/MEMS and the communication systems. Our analysis applies also to nonlinear resonators based on microwave cavities, which operate in the range $\lesssim 10$~GHz. 
\noindent

The idea of the approach is to use a driven vibrational system as a {\it bi-directional} bifurcation amplifier. Not only does such an amplifier provide a high sensitivity to the input signal, but it also allows avoiding downconversion of the signal prior to its processing. It also directly digitizes the output. The system is envisioned to operate in the following way. The input signal is phase-modulated, see Fig.~\ref{fig:hysteresis}(c). It has the form $F_S\cos(\omega_Ft+\phi)$, with the phase $\phi$ taking on discrete values $0$ and $\pi$. They correspond to the transmitted data ``0" and ``1". The signal is applied directly to the vibrational mode and is added to the same-frequency drive $F\cos\omega_Ft$. Depending on the phase $\phi$, the overall drive amplitude either increases or decreases. If the change of the drive amplitude is sufficiently strong, so that it  crosses the bifurcational value in Fig.~\ref{fig:hysteresis}(a), the mode can switch from the initially occupied state to a state with a significantly different vibration amplitude. For example, let the mode be initially in the low-amplitude state [the lower branch in Fig.~\ref{fig:hysteresis}(a)], where there arrives a signal with phase $\phi=0$. Then the driving amplitude becomes $F+F_S$. If it is larger than $F_{B2}$, the mode will switch to the high-amplitude state (the upper branch). Alternatively, if the mode was on the upper branch in Fig.~\ref{fig:hysteresis}(a) and there arrives a signal with phase $\phi$ and with the amplitude such that $F-F_S<F_{B1}$, the mode will switch to the lower branch, see also Fig.~\ref{fig:hysteresis}(b). The values of the vibration amplitudes on the upper and lower branches are significantly different, and therefore they provide a ``digital  output" of the transmitted data.

In contrast to the linear signal processing with nanomechanical systems \cite{Tadokoro2017}, the proposed approach works provided the signal amplitude $F_S$ is above some threshold value (we note that a threshold is also involved in post-processing of a linearly detected signal at the stage of digitization). The threshold is determined by the distance between the bifurcational values $F_{B1}$ and $F_{B2}$, as the overall drive amplitude  has to cross one or the other of them depending on the phase of the signal. For a nonlinear mode,  the difference  $F_{B2}-F_{B1}$ depends on the detuning of the drive frequency $\omega_F$ from the mode eigenfrequency $\omega_0$ \cite{LL_Mechanics2004}. At a critical value of $\omega_F-\omega_0$, the values $F_{B1}$ and $F_{B2}$ merge. Their dependence on $\omega_F-\omega_0$ near the critical point is sketched in Fig.~\ref{fig:hysteresis}(b). This makes the vicinity of the critical point particularly interesting for the considered bi-directional bifurcation amplifier, as it allows lowering the detection threshold.

 The limitations on the distance to the critical point are two-fold. First, the difference of the vibration amplitudes on the upper and lower branch has to be sufficiently large; this difference  sharply increases with the increasing distance to the critical point, much faster than $F_{B2}-F_{B1}$, see below. Another constraint comes from the slowing down of the system near the critical point. It limits the minimal duration of the pulse. As we show, the product of the signal amplitude times the minimal pulse duration goes down as the system approaches the critical point.

A serious constraint comes from the requirement that the signal processing is noise-protected. As we show, fluctuations sharply increase near the threshold value of the signal amplitude.  They can lead to an error when the bi-directional bifurcation amplifier switches in response to a comparatively weak signal, a noise effect that has not been explored so far, to the best of our knowledge.  We develop an approach that allows us to study the errors related to such switching. Of interest is the case where the error is small. The proposed method for evaluating the probability of a small error is close to the technique of large rare fluctuations \cite{Freidlin_book} and extends the theory of rare fluctuations in driven vibrational systems \cite{Dykman1979a,Dykman1980}. 

\section*{Results}
\noindent
{\bf The model.}
We use the standard model of vibrational dynamics of NEMS/MEMS. In this model, a nonlinear vibrational mode  is
 described as a resonantly driven Duffing oscillator \cite{Schmid2016,Dykman2012b} with equation of motion
\begin{align}
\label{eq:eom}
\ddot q + 2\Gamma\dot q + \omega_0^2 q + \gamma q^3 = F\cos \omega_Ft + \xi_0(t).
\end{align}
Here, $q$ is the oscillator coordinate, $\Gamma$ is the friction coefficient, $\omega_0$ is the eigenfrequency, and $\gamma$ is the nonlinearity parameter. We assume that, as it is usually the case in the experiment, the mode is underdamped and the nonlinearity-induced change of the vibration frequency is small, $\Gamma, |\gamma\langle q^2\rangle|/\omega_0 \ll \omega_0$.  The term $\propto F$ describes the driving force, which we assume to be close to resonance, $|\omega_F-\omega_0|\ll \omega_0$, in which case the driving can be comparatively weak to cause  bistability of forced vibrations. Function $\xi_0(t)$ in Eq.~(\ref{eq:eom}) describes noise. One can think of this noise as being a Gaussian white noise, $\langle\xi_0(t)\xi_0(0)\rangle = 2D_0\delta (t-t')$, so that the overall dynamics of the mode is Markovian. In fact, the results below apply for much less restrictive conditions where the dynamics in the laboratory frame is non-Markovian, but becomes Markovian in the rotating frame in the time slow compared to $\omega_0^{-1}$, cf. \cite{DK_review84}. In the presence of the signal $F_S\cos(\omega_Ft+\phi)$, the force amplitude $F$ is changed from the reference value in the absence of the signal $F_0$ to $F_0+F_S\cos\phi(t)$, with $\phi$ taking on the values of $0$ and $\pi$.

In the absence of the noise and the signal, after a transient process with characteristic duration $\sim \Gamma^{-1}$, the driven oscillator (\ref{eq:eom}) settles in a state of forced vibrations at frequency $\omega_F$. In such a state $q(t)=A\cos(\omega_F t+ \theta)$. When one substitutes this solution into Eq.~(\ref{eq:eom}), one obtains a cubic equation for $A$. The solution of this cubic equation is sketched in Fig.~\ref{fig:hysteresis}(a). The equation has one real positive root for $F$ outside the interval $(F_{B1},F_{B2})$ and three real roots inside this interval. The small-$A$ and large-$A$ roots correspond to coexisting stable states of forced vibrations with small and large amplitude. The intermediate root corresponds to unstable vibrations \cite{LL_Mechanics2004}.  

The idea of the bi-directional bifurcation amplifier is that, when $F_0+F_S>F_{B2}$, the system vibrates in the large-amplitude state. For $F_0-F_S<F_{B1}$ the system vibrates in the small-amplitude state. Therefore the change of the phase of the signal causes a large change of the vibration amplitude. From the above expressions, the minimal amplitude of the signal that can be detected with such an amplifier is $F_S^{\min} = (F_{B2}-F_{B1})/2$.

The bifurcational values of the amplitude $F_{B1},F_{B2}$ depend on the drive frequency, as mentioned earlier. They merge together at a critical point $K$, where the field amplitude, the frequency detuning, and the mode amplitude are
\begin{align}
\label{eq:critical_point}
F_K = 16( \Gamma^3\omega_F^3/3^{5/2}\gamma)^{1/2},\qquad \delta\omega_K = \sqrt{3}\Gamma,\qquad A_K =4(\omega_F\Gamma/3^{3/2}\gamma)^{1/2}  \qquad (\delta\omega = \omega_F-\omega_0)
\end{align}
Here we assumed $\gamma >0$; for $\gamma <0$ one should replace $\gamma \to |\gamma|$ and the detuning of the drive frequency from the oscillator eigenfrequency $\delta\omega \to -\delta\omega$. Near the critical point we have $F_{Bn}^2 -F_K^2= F_K^2\left\{(\sqrt{3}/2)\Delta_F  +[(-1)^n/2^{1/2}3^{3/4}]\Delta_F^{3/2}\right\}$, where $\Delta_F = (\delta\omega-\delta\omega_K)/\Gamma$ and $n=1,2$. This shows that the bifurcation field values diverge as $\pm\Delta_F^{3/2}$ with increasing frequency detuning, cf. Fig.~\ref{fig:hysteresis}(b). At the same time, the bifurcational values of the squared vibration amplitudes [the heights of the ``end points" of the upper and lower branches in Fig.~\ref{fig:hysteresis}(a)] diverge as $12^{-1/4}A_K^2\Delta_F^{1/2}$ (Methods). 

The difference between the bifurcational values of the vibration amplitudes  $\delta A_B$  characterizes the strength of the output signal, as it is clear from Fig.~\ref{fig:hysteresis}(a). From the above estimates we have for the minimal amplitude of the signal $F_S^{\min}$ and for $\delta A_B$ near the critical point
\begin{align}
\label{eq:signal_amplitude}
F_S^{\min}=12^{-3/4}F_K\Delta_F^{3/2}, \qquad \delta A_B = 12^{-1/4}A_K\Delta_F^{1/2}, \qquad \Delta_F =\frac{\omega_F-\omega_0}{\Gamma} -\sqrt{3}.
\end{align}   
This shows that the ``output" $\delta A_B$ increases with the increasing distance to the critical point much faster than the minimal signal amplitude. Therefore the vicinity of the critical point is a promising region for making a bi-directional bifurcation amplifier. 

\hfill

\noindent
{\bf Dynamics of the bi-directional amplifier near the critical point.} To study the operation of the bifurcation amplifier we first change from the fast oscillating variables $q(t), p(t)= \dot q(t)$ to the variables $Q(t),P(t)$ (the quadratures), which vary on the time scale much larger than the vibration period $2\pi/\omega_F$ and characterize the amplitude and phase of the vibrations. The analysis of the dynamics of these variables is simplified by the fact that, formally, the critical point of a driven oscillator is a co-dimension-2 bifurcation point \cite{Dykman1980}. Near such a point one of the variables is much slower than the other \cite{Guckenheimer1987}. This allows reducing the description to a one-variable Langevin equation in slow time $t'=\sqrt{3}\Gamma t$ for an auxiliary overdamped system with coordinate $x$, which is equal to the deviation  of the quadrature $P$ from its value at the critical point (Methods),
\begin{align}
\label{eq:potential}
\frac{dx}{dt'} = -\frac{dU}{dx} + \xi_P(t'), \qquad U(x) = \frac{1}{3\sqrt{3}}x^4 -\frac{1}{6}\Delta_F x^2 + d_Fx, \qquad d_F = \frac{1}{3\sqrt{2}}\Delta_F -\frac{2\sqrt{2}}{3\sqrt{3}}\frac{F-F_K}{F_K}
\end{align}
The variable $x$ gives the vibration amplitude near the critical point, $A\approx A_K(1+x/\sqrt{2})$. 

It follows from Eq.~(\ref{eq:potential}) that, for a given force frequency,  the optimal choice of the force amplitude in the absence of the signal is $F_0 =F_K[1+(\sqrt{3}/4)\Delta_F$]. This choice corresponds to $d_F=0$ in the absence of the signal and makes the response to the signals with the opposite phase symmetric: $d_F$ has the opposite sign but the same absolute value for $F=F_0\pm F_S$, specifically, $d_F=-(8/27)^{1/2}F_S\cos\phi/F_K$. The dependence of $F_0$ on $\delta\omega_F$ is shown by the dashed line in Fig.~\ref{fig:hysteresis}(b). For $d_F=0$, the potential $U(x)$ is a symmetric double-well potential. Its minima correspond to coexisting stable vibrational states in the absence of a signal, with amplitudes $A_K[1\pm (\sqrt{3}\,\Delta_F/8)^{1/2}]$. 

\begin{figure}[h]
\begin{center}
\includegraphics[height=3.2cm]{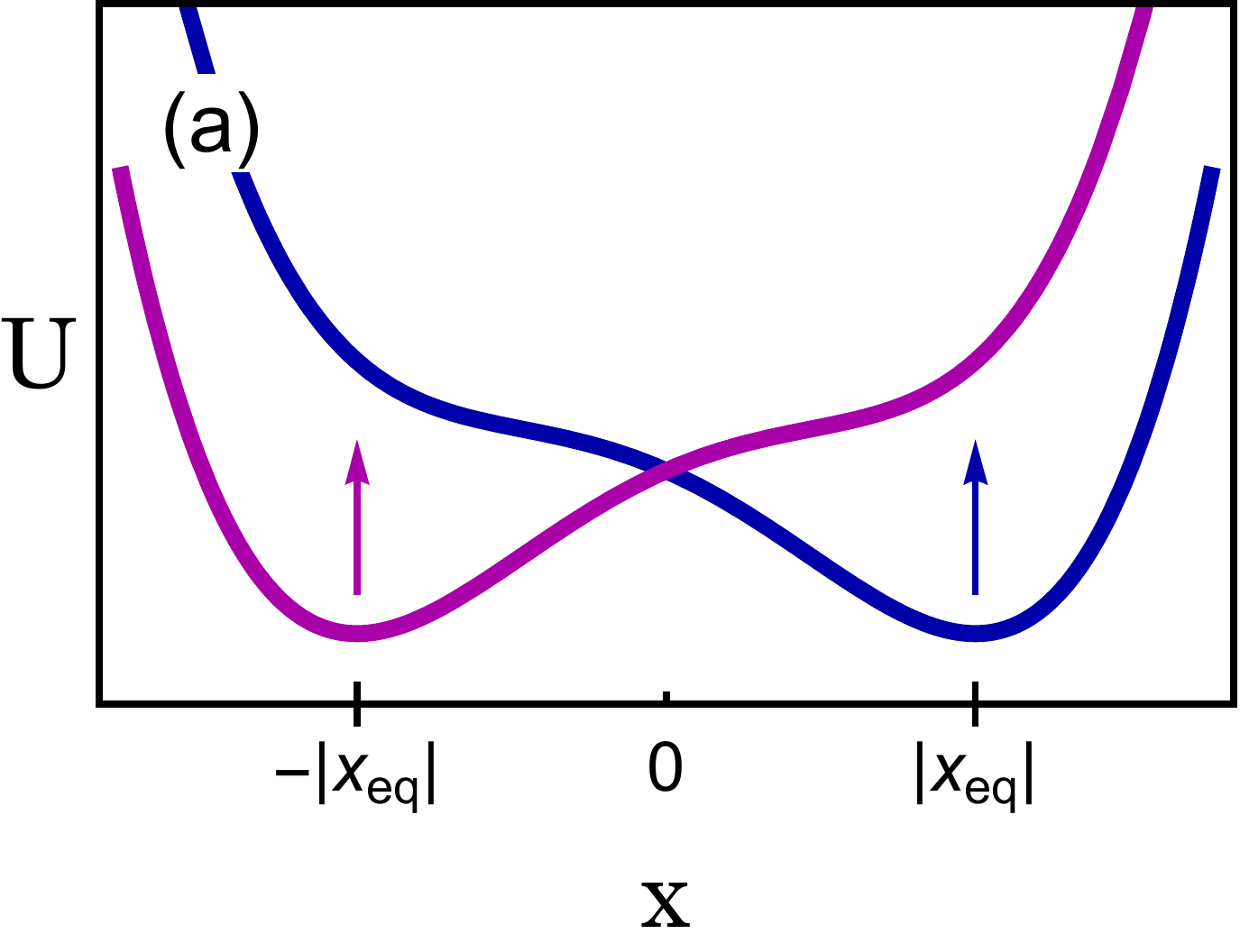} \hspace*{0.6in}
\includegraphics[height=3.2cm]{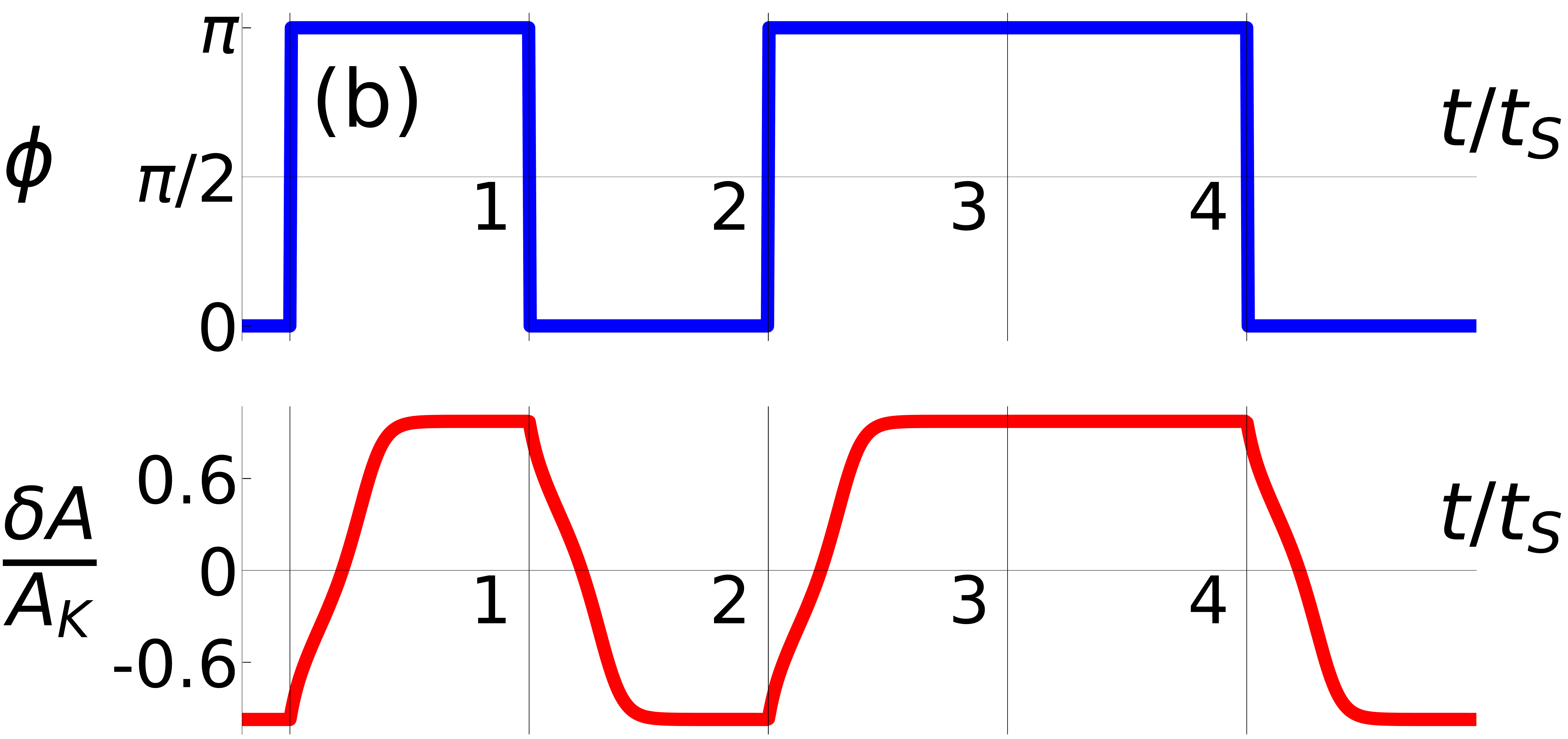}
\end{center}
\caption{{\bf Response of the bifurcation amplitifer} (a) A sketch of the effective potential for the bi-directional amplifier near the critical point, Eq.~(\ref{eq:potential}). In the presence of a signal, the potential has one minimum, which corresponds to a single state of forced vibrations of the NEMS/MEMS mode. The position of the minimum $\pm|x_{\rm eq}|$, and thus the value of the amplitude $A = A_K(1+x/\sqrt{2})$ depend on the phase of the signal $\phi$. The change of $\phi$ between 0 and $\pi$ causes the change of the potential from having the minimum at $x>0$ (blue curve) to having the minimum at $x<0$ (magenta curve). The arrows indicate how the potential changes with the change of the phase. (b) The change of the mode amplitude $\delta A = A-A_K$ in response to a change of the phase of the signal. The phase can switch between $0$ and $\pi$  at the multiples of the duration of the signal $t_S$ (upper panel). The switching of the mode is delayed, as seen in the lower panel. The parameters are $\Delta_F = 0.3\sqrt{3}$, $ t_S= 25/\sqrt{3}\Gamma$, and  $(F_S-F_S^{\min})/F_K = (3/2)^{3/2}/10 \approx 0.184$.  }
\label{fig:response}
\end{figure}

By making the coefficient $d_F$ positive or negative, the signal tilts the potential to the right or to the left, as shown in Fig.~\ref{fig:response}(a). Signal detection with the bifurcation amplifier implies that the tilted potential has only one minimum, which corresponds to the single state of forced vibrations. This condition imposes a threshold on $|d_F|$. Since $|d_F|\propto F_S$ for the chosen $F_0$, such a threshold determines the minimal required signal amplitude $F_S^{\min}$. The expression for $F_S^{\min}$ that follows from Eq.~(\ref{eq:potential}) coincides with Eq.~(\ref{eq:signal_amplitude}). 

The amplifier operates in the following way. In the presence of a signal with a given phase, the potential $U(x)$ is monostable, and in the absence of noise the auxiliary system (\ref{eq:potential}) goes to its minimum. Respectively, the mode goes to the state with the corresponding amplitude. For $\phi=0$, the minimum of $U(x)$ is at $|x_{\rm eq}|>0$ [the real root of $U'(x)=0$ for $\phi=0$] and the amplitude of the vibrational state $A$ is larger than $A_K$. For $\phi=\pi$ the minimum $U(x)$ is at $x=-|x_{\rm eq}|$ and the amplitude is smaller than $A_K$. The value of  $|x_{\rm eq}|$ and thus the change of the amplitude depend on $F_S$ nonlinearly, with $|x_{\rm eq}|\propto F_S^{1/3}$ for $F_S/F_S^{\min}\gg 1$.  When the phase of the drive changes and the potential switches, cf. Fig.~\ref{fig:response}(a), the initially occupied minimum of $U(x)$ disappears and the auxiliary system moves to the new minimum. Respectively, the mode amplitude changes to the other value. The kinetics of this switching is described by Eq.~(\ref{eq:potential}) and is illustrated in Fig.~\ref{fig:response}(b).

An important characteristic is the time it takes to approach the new vibration amplitude when the phase of the signal changes. From Eq.~(\ref{eq:potential}), in the absence of noise and for sufficiently strong signal, $F_S/F_S^{\min}\gg 1$, this time is determined by the relaxation time $t_r\propto 1/U''(|x_{\rm eq}|)$ of the system near the equilibrium position. From Eq.~(\ref{eq:potential}), $ t_r \sim (F_S/F_K)^{-2/3}/\Gamma$. Note that in this case $\Gamma t_r \ll 1/\Delta_F$. Given that $F_S^{\min}\propto \Delta_F^{3/2}$ and that $t_r$ imposes the lower bound on the signal duration $t_S$, this inequality shows that, by approaching the critical point, one can reduce the product $F_St_S$. This is an important advantageous feature of the considered bi-directional bifurcation amplifier. 

\hfill

\noindent
{\bf Universality near the detection threshold.} Switching in response to a signal acquires universal features in the critical region where the signal amplitude approaches the threshold value, $F_S-F_S^{\min}\ll F_S^{\min}$. Understanding the dynamics near the threshold is important for applications of the bifurcation amplifier. Of particular importance is to estimate how the required duration of the pulse changes near the threshold.

For $F_S=F_S^{\min}$, along with the minimum, the potential $U(x)$ has another stationary point, which is an  inflection point $x_{\rm infl}$. Its sign is determined by the sign of $F_S\cos\phi$ [we remind that in the expression for the potential (\ref{eq:potential}) $d_F = -(8/27)^{1/2}F_S\cos\phi/F_K$]. For $\phi=0$, we have $x_{\rm infl} =- |x_{\rm infl}|$ with $|x_{\rm infl}|=(\Delta_F^{1/2}/4\sqrt{3})^{1/2}$ ($x_{\rm infl} = |x_{\rm infl}|$ for $\phi=\pi$). Since $-|x_{\rm infl}|$ is a stationary state of the system in the absence of noise for $F_S=F_S^{\min}$, for $F_S$ close to $F_S^{\min}$ the motion near the inflection point is slow. It is this slowness that leads to universal features of switching between the states due to a weak signal. It also leads to comparatively strong fluctuations. The effect of noise here is qualitatively different from the noise effects familiar in other problems of nano- and micromechanics and requires new means to be understood.

We note that the signal-induced change of the vibration amplitude $A$ does not go to zero for $F_S\to F_S^{\min}$, but rather $A-A_K \approx \sqrt{2} A_K|x_{\rm infl}|\cos\phi \equiv \pm \sqrt{2} A_K|x_{\rm infl}|$. This is another advantageous feature of the bifurcation amplifier.  

To analyze the dynamics near the threshold, we introduce the deviation of the signal amplitude from its threshold value $\delta F_S = F_S - F_S^{\min}\ll F_S^{\min}$ and expand the potential $U(x)$ about $-|x_{\rm infl}|$ keeping the lowest-order relevant terms in $x+|x_{\rm infl}|$. We can then write Eq.~(\ref{eq:potential}) for the scaled variable $ X\propto x+|x_{\rm infl}|$ in the scaled time $\tau$ as 
\begin{align}
\label{eq:critical_dynamics}
&\frac{dX}{d\tau} = -\frac{dV}{dX} + \xi(\tau), \qquad V(X)=-\frac{1}{3}X^3 - X, \qquad 
\langle \xi(\tau)\xi(0)\rangle = 2D\delta(\tau ).
\end{align}
Here, 
\begin{align}
\label{eq:scaled_variables}
X= a_X\left(\frac{\delta F_S}{F_K}\right)^{-1/2}(x+|x_{\rm infl}|), \qquad
\tau =a_\tau \left(\frac{\delta F_S}{F_K}\right)^{1/2}\Gamma t,
\qquad  D = a_D\left(\frac{\delta F_S}{F_K}\right)^{-3/2}D_0
\end{align}
with $a_X = 3^{3/8}(\Delta_F/2)^{1/4}$, $a_\tau = (32\Delta_F)^{1/4}/3^{5/8}$, and $a_D= 3^{15/8}\gamma (2\Delta_F)^{1/4}/64\omega_F^3\Gamma^2$. As seen from Eq.~(\ref{eq:scaled_variables}), the variable $X$ is the variable $x$ stretched about $-|x_{\rm infl}|$ by the large factor $\propto (\delta F_S)^{-1/2}$. The dimensionless time $\tau$ is  slow compared to $t$, $d\tau/dt\propto (\delta F_S)^{1/2}$.  Importantly, the noise intensity $\tilde D$ is increased compared to the intensity of the noise $D_0$ that drives the mode [cf. Eq.~(\ref{eq:eom})] by the factor $\propto (\delta F_S)^{-3/2}$. This scaling reflects the fact that the dynamics of the mode is slowed down if the potential $U(x)$ is almost flat, as it is the case for small $\delta F_S$ and $x$ near $-|x_{\rm infl}|$. Equations (\ref{eq:critical_dynamics}) and (\ref{eq:scaled_variables})  show that of interest is the region of $|x+x_{\rm infl}|\ll 1$, where $|X|\lesssim 1$. This allowed us to disregard in $V$ the term $\propto (\delta F_S/F_K)^{1/2}X^4$.

In the absence of noise, function $X(\tau)$ has a simple form
\begin{align}
\label{eq:deterministic}
X_{\rm det}(\tau) = -\cot\tau,
\end{align}
where we have chosen $X_{\rm det}\to -\infty$ for $\tau\to +0$. Equation (\ref{eq:deterministic}) describes the trajectory along which the mode switches in response to the change of the signal phase $\phi$ from $\pi$ to 0. Indeed, for $\phi=\pi$ the vibration amplitude of the mode corresponds to the equilibrium position $-|x_{\rm eq}|$ of the auxiliary system (\ref{eq:potential}).  From Eq.~({\ref{eq:scaled_variables}), this position corresponds to a very large negative value of the scaled variable $X$,  $X\approx -a_X(\delta F_S/F_K)^{-1/2}|x_{\rm infl}|$. To the leading order in $\delta F_S$, one can set $X\to -\infty$ for $x=-|x_{\rm eq}|$. In the switching, the auxiliary system moves from $-|x_{\rm eq}|$ to $|x_{\rm eq}|$, and respectively, for $x=|x_{\rm eq}|$ one can set $X\to \infty$. It follows from Eq.~(\ref{eq:deterministic}) that, in the dimensionless time $\tau$, the switching duration is $\tau_{\rm sw}=\pi$. In the dimensional time, it is $t_{\rm sw}= \pi(\delta F_S/F_K)^{-1/2}/\Gamma a_\tau$  [here we disregarded the contribution of the relaxation time $t_r\sim (\Gamma\Delta_F)^{-1}$, which does not contain the large factor $ (\delta F_S/F_K)^{-1/2}$]. The long switching duration imposes an important constraint on the minimal duration of the signal, which should exceed a threshold value determined by the signal amplitude, $t_S>t_{\rm sw}\propto |\delta F_S|^{-1/2}$.

The actual trajectory followed by the system differs from $X_{\rm det}(\tau)$ because of the noise. This is shown in Fig.~\ref{fig:trajectory}(a).  As seen from Fig.~\ref{fig:trajectory}(a), the noise can accelerate transversing the region $|X|\lesssim 1$, making the switching faster and thus improving the performance of the bifurcation amplifier. However, it can also slow the system down. 

\hfill

\noindent
{\bf Error probability as a rare event problem.} If the noise is weak, most likely the vibrational mode will switch over the time  $t_{\rm sw}= \pi(\delta F_S/F_K)^{-1/2}/\Gamma a_\tau$. The probability of having not switched is the error probability. It is given by the probability for the auxiliary system (\ref{eq:potential}) to have not approached  the equilibrium position $|x_{\rm eq}|\approx 2|x_{\rm infl}|$. In terms of the $X$-variable, this corresponds to being in the region $X < 3a_X(\delta F_S/F_K)^{-1/2}|x_{\rm infl}|$. The right-hand side of this inequality is a very large number. For the considered switching from the state at $X\to -\infty$, it is convenient to introduce the probability $P_{\rm err}(X,\tau)$ to be in the region to the left of $X$ at time $\tau$; having not switched by the time $\tau$ is given by $P_{\rm err}(X,\tau)$ for large $X$. On the other hand, if the system's coordinate is larger than $X\gg 1$, the noise can be disregarded, and the system will go to $X\to\infty$ over the scaled time $1/X\ll 1$, as seen from Eq.~(\ref{eq:critical_dynamics}). Function $P_{\rm err}(X,\tau)$ can be conveniently expressed in terms of the probability density of the auxiliary system $\rho(X,\tau)$ to be at point $X$ at time $\tau$ as   $P_{\rm err}(X,\tau)=\int\nolimits_{-\infty}^X\rho(X',\tau)dX'$. 


If the system has not switched, it means that the noise, which is very weak on average, could efficiently compensate the force $-dV/dX$ in Eq.~(\ref{eq:critical_dynamics}) for a sufficiently long time. This is a large rare fluctuation. Different specific realizations of the noise can lead to non-switching, but for a Gaussian noise all of them have exponentially small probability densities, which are also exponentially different. A natural approach to the problem of the error probability then is to find the most probable appropriate realization of the noise out of all the improbable realizations. Such optimal noise drives the auxiliary system along a certain trajectory \cite{FeynmanQM,Dykman1979a}. A rigorous formulation for finding this trajectory has been developed in the mathematical literature \cite{Freidlin_book}. However, to the best of our knowledge,  the problem we are discussing has not been considered and the approach described below has not been developed.

The physical picture of the dynamics of the noisy auxiliary system  is as follows. By the time $\tau \gtrsim \pi/2$ the system will have left the area of large negative $X$ where the motion is fast and the effect of the noise is negligible, cf. Fig.~\ref{fig:trajectory}(a).  The motion slows down in the range $|X|\lesssim 1$ and the effect of the noise accumulates here. As seen from Fig.~\ref{fig:trajectory}(a), the noise can accelerate the passage through the range $|X|\lesssim 1$, in which case the system switches faster than in the absence of the noise. However, the noise can also slow the system down, delaying the switching. After the system goes through the range $|X|\lesssim 1$ it again moves fast. For $X\gg 1$ the noise can be largely disregarded. The distribution $\rho(X,\tau)$ for $X\gg 1$ corresponds to the flux (current) toward $X\to \infty$. This current is ultimately determined by the distribution for $X\lesssim 1$. The diffusion component of the current is small. 

For weak noise the probability to stay in the region $|X|\lesssim 1$ for time $\tau \gtrsim \pi$ is exponentially small. Therefore $\rho(X,\tau)$ is also exponentially small, as is the overall error probability $P_{\rm err}(X,\tau)$. Of primary interest is then the exponential factor in the expression for $\rho(X,\tau)$.  In the considered case of a small noise intensity $D$,  this factor can be found by solving the Fokker-Planck equation for $\rho(X,\tau)$ in the eikonal approximation \cite{Stratonovich1971,Freidlin_book,Graham1973,Ludwig1975}. In this approximation, the problem is reduced to the problem of the Hamiltonian dynamics of a classical mechanical particle with coordinate $X$, momentum $\Pi$, the Hamiltonian ${\cal H}(X,\Pi)$, and the mechanical action $S(X,\tau)$ (Methods),
\begin{align}
\label{eq:eikonal}
\rho(X,\tau) =C_{X,\tau} \exp[-S(X,\tau)/ D], \qquad S(X,\tau)= -E\tau + \int_{-\infty}^X \Pi(X')dX',
\qquad {\cal H}(X,\Pi) = \Pi^2 -\Pi \frac{dV}{dX}.
\end{align}
Here, $E$ is the value of the energy with which the particle is moving along the Hamiltonian trajectory that starts at $\tau=0$ at $X\to -\infty$ and arrives to $X$ at time $\tau$. This is  the most probable trajectory followed by the auxiliariy system (\ref{eq:critical_dynamics}) as it moves from $X\to -\infty$  to point $X$.  Equation (\ref{eq:eikonal}) applies where the ratio $S(X,\tau)/D$ is large. Respectively, the probability density $\rho(X,\tau)$ is exponentially small, as expected. The prefactor $C_{X,\tau}$ is of order 1, it is independent of the noise intensity $D$. Of primary interest therefore is the action $S(X,\tau)$. 

\begin{figure}[h]
\begin{center}
\includegraphics[height=2.cm]{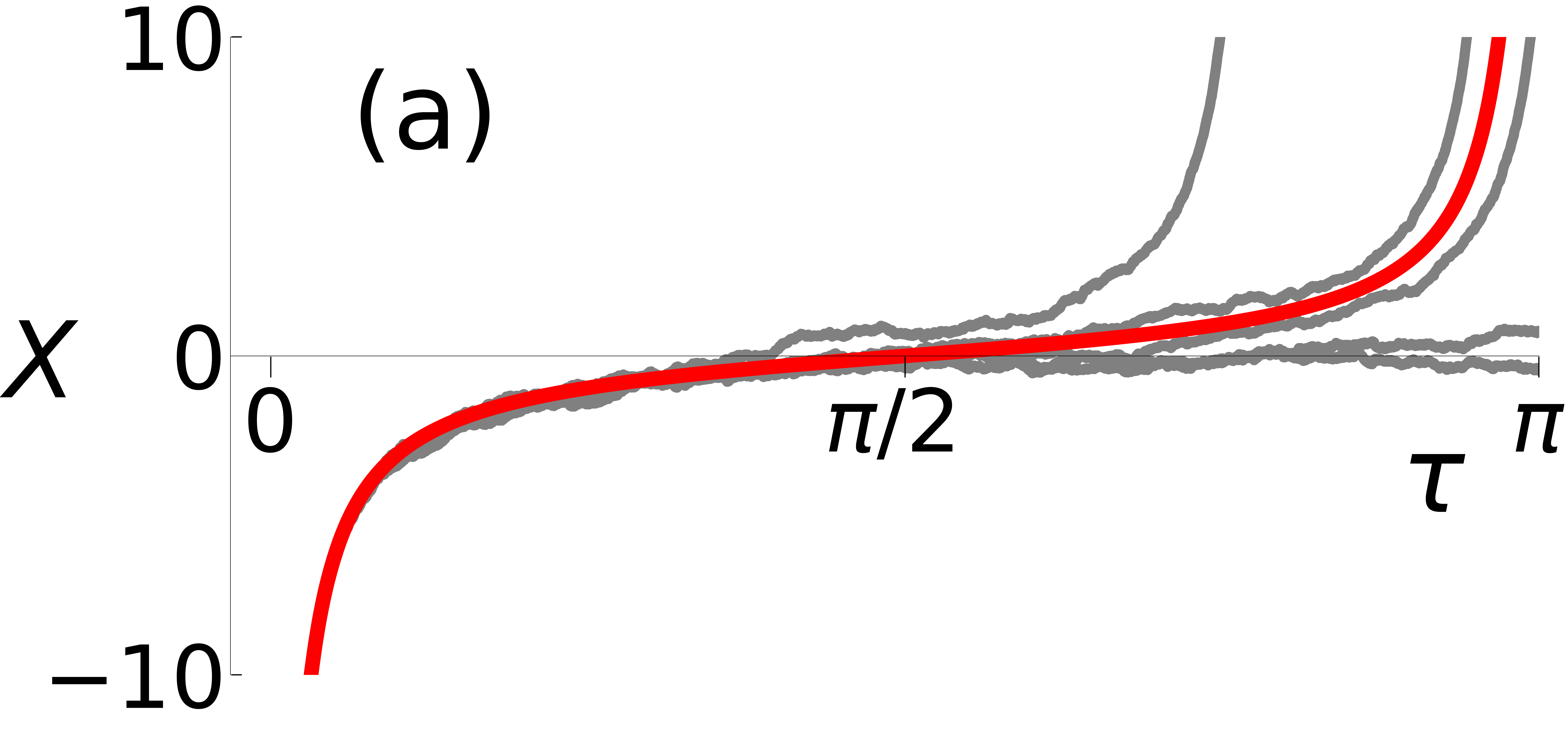} \hspace*{0.2in}
\includegraphics[height=2.5cm]{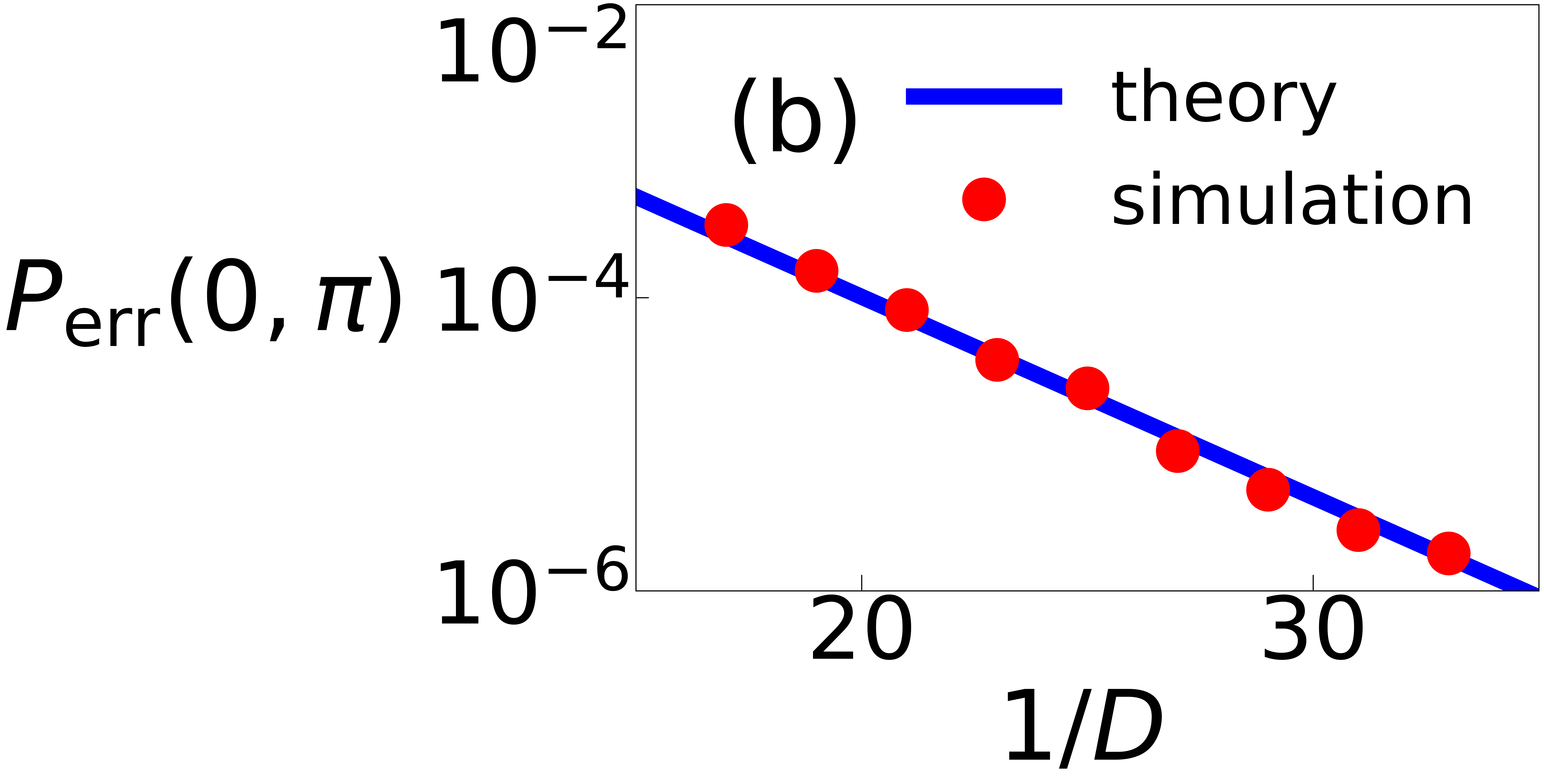}\hspace*{0.2in}
\includegraphics[height=2.5cm]{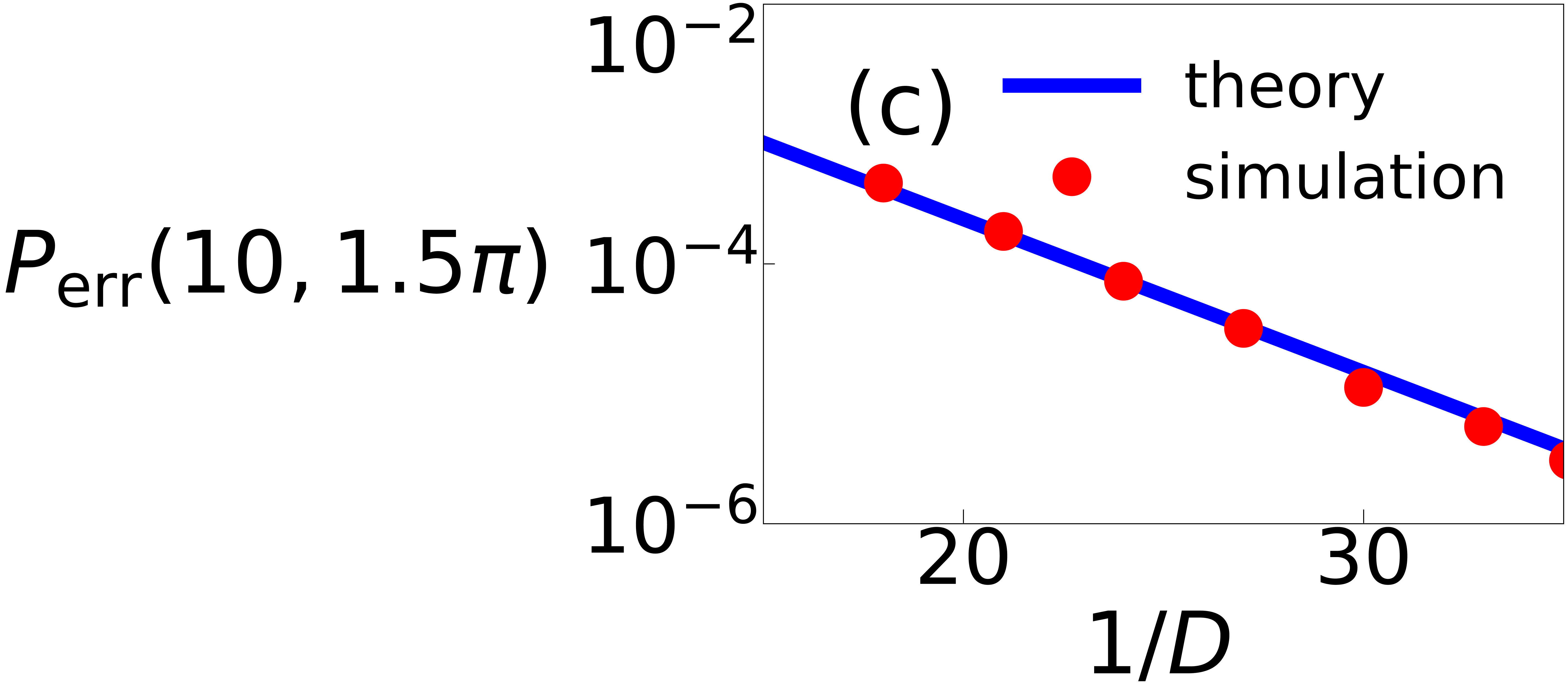}\hspace*{0.2in}
\end{center}
\caption{{\bf Switching errors near the threshold signal amplitude} (a) The trajectories followed by the auxiliary system in switching. The bold solid red line is the noise-free trajectory, Eq.~(\ref{eq:deterministic}). The actual trajectories in the presence of noise deviate from this trajectory primarily because of the slowing down of motion that occurs in the region $|X|\lesssim 1$ or, equivalently, for $\tau$ close to $\pi/2$. The figure shows several realizations of the noisy trajectories obtained by simulating the Langevin equation of motion (\ref{eq:critical_dynamics}) for the noise intensity $D=0.15$. (b) The activation dependence of the error probability calculated as the probability to stay in the region $X\leq 0$ for the dimensionless time $\tau=\pi$. The data points are the results of the simulations. The slope of $\ln P_{\rm err}$ is given by Eqs.~(\ref{eq:FPE}) and (\ref{eq:action_explicit}).  (c) The same as in (b) for $X \leq 10$ and $\tau=1.5 \pi$. From Eq.~(\ref{eq:action_explicit}), the value of $D\ln P_{\rm err}$ in this case weakly depends on $X$ and gives the full error probability for $X<\infty$.   }
\label{fig:trajectory}
\end{figure}

On physical grounds it is clear that, as we increase $\tau$, the auxiliary system will be more and more likely to switch. To minimize the noise-induced error we have to increase the switching time compared to the noise-free value $\tau_{\rm sw}=\pi$. As we show (Methods), $S(X\tau)$ becomes large for $X\gg 1$, thus reducing the error probability, already for $\tau=1.5\pi$. Such value of $\tau$ is not a hard limit: by reducing the noise intensity one can make the error rate small for smaller time. However, our numerical results refer to this value of time. 

An important condition of the applicability of Eq.~(\ref{eq:eikonal}) is  $D|\Pi(X)| \gg |d^2V/dX^2|$. In the problem at hand $\Pi(X)<0$ and $|\Pi(X)|$ decreases with increasing $X$ for $X\gg 1$, whereas $|d^2V/dX^2|$ increases (Methods). For sufficiently  large $X$ the eikonal approximation no longer applies and the distribution $\rho(X,\tau)$ goes over into that of the noise-free current. The exponent of the error rate $S_{\rm err}(\tau)/D$ is given by $S(X,\tau)/D$ for such $X$. It weakly depends on $X$. This picture makes the problem at hand different from the problems usually studied in the theory of rare events. The explicit form of the action $S_{\rm err}(\tau)$ is given by Eq.~(\ref{eq:action_explicit}) in Methods. The leading-order term in $S_{\rm err}$ is $(\tau/4)- (2^{3/2}/3)$. For the correction to be small it is necessary that $64\exp(-2^{1/2}\tau)\ll 1$. For the value $\tau = 1.5\pi$ used in our numerical simulations the left-hand side of this inequality is $\approx 0.08$. 

In Fig.~\ref{fig:trajectory}(b) and (c) we compare the slope of the error probability $P_{\rm err}(X,\tau)$ with the asymptotic expressions based on the eikonal approximation. The data was obtained by numerically simulating the noisy dynamical system (\ref{eq:critical_dynamics}) and counting the portion of the trajectories that ended at points $X'<X$ at time $\tau$. 
It is seen that, for already moderately small noise intensity, the error probability becomes small. As expected, it displays an activation-type dependence on the noise intensity $D$. The slope of the exponent approaches the asymptotic value $S(X,\tau)/D$ for small $D$. 

\section*{Discussion}

Since nano- and micromechanical systems are small, their vibrations become nonlinear already for small vibration amplitude. This paper shows that the vibration nonlinearity can be used to develop a new type of signal detectors. A basic  consequence of the nonlinearity is the bistability of the vibrations excited by a resonant force. The decay rates of nano-mechanical systems are small, and therefore the bistability emerges already for comparatively small force amplitude. It exists in a certain range of the force amplitudes, which is limited both from above and from below. When a signal, depending on its phase, increases or decreases the driving amplitude, the overall force changes, respectively, to the values above or below the bistability region. The values of the vibration amplitude on the opposite sides of the bistability region differ strongly. Therefore the change of the phase of the signal leads to a large change of the vibration amplitude. 

A convenient region for the operation of the amplifier is the vicinity of the critical point of the nonlinear vibrational mode. At this point the bifurcational values of the driving amplitude merge. Near the critical point the threshold for the signal amplitude $F_S^{\min}$ is  small. It varies with the distance to the critical point as $\Delta_F^{3/2}$, where $\Delta_F$ is the scaled difference between the driving force frequency and its value at the critical point. In contrast, the minimal change of the amplitude of the forced vibrations scales as $\Delta_F^{1/2}$, i.e., in the appropriate units it is much larger than the threshold signal amplitude for small $\Delta_F$. 

The dynamics of the system acquires universal features for  the signal amplitude $F_S$ close to $F_S^{\min}$. The minimal signal duration scales as $(F_S-F_S^{\min})^{-1/2}$. Importantly, the system becomes sensitive to  noise. We introduce the parameter that describes the probability of an error.  The error of interest is that the mode has not switched in response to the changed phase of the signal over a given time, which is determined by the signal duration. We develop a technique to find the exponential factor that gives the error probability for small noise intensity. As we show, this probability displays an activation-type dependence on the noise intensity, and we calculate the exponent. The results of numerical simulations of the mode dynamics are in good agreement with the theoretical predictions.


\section*{Methods}

{\bf Slow variables.} The dynamics of underdamped weakly nonlinear oscillators is conveniently analyzed using the method of averaging \cite{Kryloff1947,Nayfeh2004}. In this method, one changes from the fast oscillating coordinate and momentum to the slow variables, which describe the change of the vibration amplitude and phase in time. In addition, it is convenient to scale the variables and thus to introduce the relevant combinations of the parameters. A convenient transformation reads $q=C_{\rm res}(Q\cos\omega_Ft + P\sin\omega_Ft), \dot q = -\omega_FC_{\rm res}(Q\sin\omega_Ft - P\cos\omega_Ft)$ with $C_{\rm res} = (8\omega_F\delta\omega/3\gamma)^{1/2}$. Changing to the dimensionless time $t'=t(\delta\omega)$, we obtain from Eq.~(\ref{eq:eom}) the following equations of motion for $Q,P$:
\begin{align}
\label{eq:eom_QP}
&\frac{dQ}{dt'} = -\kappa Q+\frac{\partial g}{\partial P} + \xi_Q(t'), \quad \frac{dP}{dt'} = -\kappa P-\frac{\partial g}{\partial Q} + \xi_P(t'),\qquad 
\kappa = \Gamma/\delta\omega, \nonumber\\
&g(Q,P) = \frac{1}{4}(Q^2+P^2-1)^2 -\beta^{1/2}Q, 
\qquad \beta = 3\gamma F^2/32\omega_F^3(\delta\omega)^3
\end{align}
Here $\xi_Q(t')$ and $\xi_P(t')$ are independent white noises with intensity $D'=3\gamma D_0/16\omega_F^3(\delta\omega)^2$. In the spirit of the method of averaging, we have disregarded fast-oscillating terms in Eq.~(\ref{eq:eom_QP}). We note that this approximation corresponds to the well-known rotating wave approximation in quantum optics \cite{Mandel1995}.

In the absence of noise, the dynamics (\ref{eq:eom_QP}) is determined by two dimensionless parameters $\kappa$ and $\beta$, which characterize the relative detuning of the drive frequency from the oscillator eigenfrequency and the relative strength of the drive. One can find the stationary states by setting $dQ/dt'=dP/dt'=0$, which leads to a cubic equation for $Q^2+P^2$. At the critical point three solutions of this equation merge together. This condition gives the critical parameter values $\kappa_K = 1/\sqrt{3}, \beta_K=8/27$, whereas $Q_K=-1/\sqrt{6}$ and $P_K=1/\sqrt{2}$. If one linearizes Eqs.~(\ref{eq:eom_QP}) about the critical point, one finds that the equation for $dP/dt'$ does not contain terms that are linear in $P-P_K,Q-Q_K$. This means that $P$ is a slow variable near the critical point, whereas $Q(t')$ follows  $P(t')$ adiabatically \cite{Dykman1980}. Over the dimensionless time $t'\sim 1$, the difference $Q-Q_K$ approaches the value $(P-P_K)/\sqrt{3} -(\kappa - \kappa_K)/2\sqrt{2} +(4/\sqrt{6})(P-P_K)^2$. Substituting this expression into the equation for $P-P_K$ and taking into account that fluctuations of  $Q$ are much smaller than those of $P$, we obtain the equation for the evolution of $P(t')$, which reads $dP/dt' = -dU((P-P_K))/dP + \xi_P(t')$; the effective potential $U(x)$ is defined in Eq.~(\ref{eq:potential}). The squared vibration amplitude near the critical point, to the leading order, is $A^2 = C_{\rm res}^2(P^2 + Q^2)\approx A_K^2[1+\sqrt{2}(P-P_K)]$. We used this expression in the main text to relate the change of the vibration amplitude to the dynamics of the slow variable.

\hfill 

\noindent {\bf The eikonal approximation near the detection threshold.}  The probability of a noise-induced error near the detection threshold can be conveniently found by changing from the Langevin equation of motion of the auxiliary system (\ref{eq:critical_dynamics}) to the Fokker-Planck equation (FPE) for the probability distribution of this system $\rho(X,\tau)$ and seeking the solution of the FPE in the eikonal form \cite{Freidlin_book},
\begin{align}
\label{eq:FPE}
\partial_\tau\rho = \partial_X[(dV/dX)\rho ] + {D}\partial_X^2\rho, \qquad \rho\equiv \rho(X,\tau) = \exp[-S(X,\tau)/D].
\end{align}
Substituting the expression $\rho\propto \exp(-S/D)$ into the FPE and keeping the leading-order terms in $D$, we obtain an equation for $S(X,\tau)$ of the form $\partial_\tau S = -{\cal H}(X,\partial_X S)$, where ${\cal H}$ is given in Eq.~(\ref{eq:eikonal}). This is the Hamilton-Jacobi equation for the mechanical action $S$ of a particle with the Hamiltonian ${\cal H}(X,\Pi)$; the momentum of the particle is $\Pi=\partial_X S$ \cite{LL_Mechanics2004}. 

The Hamiltonian trajectory of the particle starts at $X\to -\infty$ for $\tau =0$: this is where the auxiliary fluctuating system was initially prepared. The region $X\to -\infty$ corresponds to the vibrational state of the mode for the phase of the signal $\phi = \pi$; at $\tau=0$  the phase changes to $\phi =0$ and the mode switches. Respectively, the auxiliary  starts moving to a new stable state.  This explains the limits of the integral in the expression (\ref{eq:eikonal}) for the action and why $S(X,0)= 0$ for $X\to -\infty$.

From the explicit form of the Hamiltonian (\ref{eq:eikonal}), we have on the Hamiltonian trajectory
\begin{align}
\label{eq:trajectory}
\frac{dX}{d\tau}= [(X^2+1)^2 +4E]^{1/2}, \qquad \Pi(X,\tau)= \frac{1}{2}\{-(X^2+1) +  [(X^2+1)^2 +4E]^{1/2}\}.
\end{align}
Function $X(\tau)$ can be expressed in terms of the Jacobi elliptic functions. However, the expression is somewhat cumbersome, and we will not use it. 

An important parameter of the trajectory is the particle energy $E$. It is given by the expression $\int_{-\infty}^X dX'[(X^2+1)^2 +4E]^{-1/2} =\tau$, which follows from Eq.~(\ref{eq:trajectory}). Formally, from Eq.~(\ref{eq:eikonal}) $\partial S/\partial\tau = -E$, and since the distribution $\rho(X,\tau)$ should decrease with increasing $\tau$, for the relevant Hamiltonian trajectories $E<0$. A significant consequence of the last inequality is that $\Pi(X,\tau)<0$. It means that $S(X,\tau)$ decreases with the increasing $X$, i.e., the distribution $\rho(X,\tau)$ increases with $X$ where the eikonal approximation applies. For large $X$ we have $\Pi(X)\approx 2E/X^2$, whereas $|d^2V/dX^2|= 2X$. Therefore the breakdown of the eikonal approximation for $|\Pi(X)|\sim D|d^2V/dX^2|$ occurs for $X\sim X_l= (|E|/D)^{1/3}$.

The energy can be found in the explicit form in the important case where $\tau$ is comparatively large. From Eq.~(\ref{eq:trajectory})
\begin{align}
\label{eq:energy}
E =-1/4+\ep, \qquad \ep \approx 16\exp(-2^{3/2}\tau) \; {\rm for}\; X=0; \qquad \ep \approx 16\exp[-2^{1/2}(\tau+X^{-1})]\;{\rm for}\; X\gg 1.
\end{align}
Of utmost interest is the case $\ep\ll 1$. In this case, using Eq.~(\ref{eq:eikonal}), one can calculate the action $S(X,\tau)$ in the explicit form. To the first order in $\ep$
\begin{align}
\label{eq:action_explicit}
&S(0,\tau) \approx  \frac{\tau}{4} - \frac{2^{1/2}}{3} +\frac{\ep}{2\sqrt{2}}; \quad
S(X,\tau) \approx  S_{\rm err}(\tau) +\frac{1-4\ep_\infty}{4X}\; {\rm for}\; X\gg 1; \quad S_{\rm err}(\tau)=
\frac{\tau}{4} - \frac{2^{3/2}}{3}+ \frac{\ep_\infty}{\sqrt{2}}.
\end{align}
Here, $\ep_\infty$ is given by Eq.~(\ref{eq:energy}) for $\ep$ in which one sets $X\to \infty$. As explained in the main text, $S_{\rm err}(\tau)$ gives the exponent of the error rate.



\newpage

\noindent
{\bf Acknowledgments}

\noindent
This work was partly supported by JSPS KAKENHI (grant JP17K14690). MID acknowledges partial support from the US National Science Foundation (DMR-1514591 and CMMI-1661618)

\hfill

\noindent
{\bf Author Contributions}

\noindent
The idea of using driven nonlinear nanomechaical resonators for signal detection was developed by the authors together. MID developed the theory of a bi-directional bifurcation amplifier. YT ran numerical simulations. All authors co-wrote the paper.

\hfill

\noindent
{\bf Competing interests:} The authors declare that they have no competing interests.

\end{document}